\documentclass[twocolumn,aps,showpacs,prl,amsmath,amssymb,floatfix,superscriptaddress]{revtex4}
\usepackage{color}
\usepackage{graphicx}% Include figure files
\usepackage{dcolumn}% Align table columns on decimal point
\usepackage{bm}% bold math
\usepackage{array}
\usepackage{float}
\usepackage{longtable}
\usepackage{mathrsfs}
\usepackage{txfonts}

\begin{document}

\title{A Renormalization-Group Study of the Symmetry-Breaking Order Parameters in Spin-Orbit Coupled Iridates and Related Systems}

\author{Tanmoy Das}\affiliation{Theoretical Division, Los Alamos National Laboratory, Los Alamos, New Mexico  87545, USA}
\author{Armin Rahmani}\affiliation{Theoretical Division, T-4 and CNLS, Los Alamos National Laboratory, Los Alamos, New Mexico  87545, USA}
%\author{Andrey Chubukov}\affiliation{Department of Physics, University of Wisconsin, Madison, Wisconsin 53706, USA}

\date{\today}


\begin{abstract}
We study the competition between various forms of density-wave-like order parameters that may arise in multi-orbital correlated materials with strong spin-orbit coupling (SOC) within a renormalization-group (RG) approach. The calculations are restricted to models with two spin-orbit split bands having strong inter-band Fermi surface nesting. We find that for such Fermi surface topology, the interplay between the inter-band nesting and SOC strongly enhances the inter-orbital Coulomb interaction (relative to other interactions) as the system approaches a stable fixed point. This results in an exotic spin-orbit density wave (SODW) to become the energetically favorable symmetry-broken state. While the conclusions are generic to such a Fermi surface topology, the band structure and the numerical results are presented for the iridate systems. We also find that the electronic fingerprints of the SODW are in better agreement with various experimental results than a conventional spin density wave (SDW), explaining the long-standing problem of the origin of the metal-insulator transition in these systems.  
\end{abstract}

\maketitle

Interaction-driven  Fermi surface (FS) instabilities can lead to spontaneous symmetry breaking and the emergence of ordered phases such as superconductivity, charge, spin, or orbital ordering. There has been significant recent interest in materials with strong spin-orbit interaction, e.g., topological insulators, noncentrosymmetric superconductors, heavy-fermions and several oxide interfaces, which exhibit unconventional phases of matter~\cite{ExoticPhaseSOC}.  The interplay of spin-orbit coupling (SOC) and Coulomb interactions can also give rise to novel exotic topological phases~\cite{topological,review}. What are the effect of SOC on FS instabilities?  SOC may favor symmetry-broken ordered states characterized by strong entangling of spin and orbital degrees of freedom, referred to spin-orbital density wave (SODW) \cite{SODWSR,SODWPRL}. Renormalizationgroup (RG) is a powerful method for analyzing the competing order parameters. This approach is applied extensively to bilayer graphene \cite{grapheneRG}, iron-based superconductors \cite{pnictideRG}, and related materials \cite{othermaterialsRG}. The objective of this paper is to provide the RG analysis of the competition between various Fermi-surface instabilities toward density-wave order in the presence of strong SOC.

We focus on instabilities that lead to a metal-insulator transition (MIT) by gapping out the FS. Traditionally, systems with narrow bandwidth in comparison with their correlation strength are prone to MIT. The recent discovery of MIT in pyroclore and perovskite iridates family \cite{NaIrOdiscovery,SrIrOdiscovery08,SrIrOdiscovery09} with 5$d$ orbitals came as a surprise since their noninteracting bandwidth, $W$, is estimated to be much larger than Coulomb interaction $U$. Many theoretical approaches including strong \cite{strongcoupling} and weak coupling \cite{weakcoupling,LDAU,TB} theories suggest that the SOC of the Ir atoms is responsible for this phenomena \cite{SrIrOdiscovery09}. The SOC splitting of the $t_{2g}$ orbitals leads to half-filled effective $J_{\rm eff}=1/2$ orbitals to be near the Fermi level with an effective bandwidth $W_{\rm eff}$ which is smaller than the interaction strength $U$. Experimental data in Sr-based iridates suggest that at the temperature $T^*\sim$70~K \cite{neutronSr327}, where the MIT sets in, the DC magnetization begins to drop, in contract to what is expected from a typical spin-ordered state. This result (and other evidence as discussed in Refs.~\onlinecite{neutronSr327,Nd227,muon,MGe,ARPES,STM}) suggests that insulators arising from such SOC-induced MIT may have  a complex form of spin-orbital order. 

\begin{figure}
%\hspace{-2.2cm}
\rotatebox[origin=c]{0}{\includegraphics[width=0.9\columnwidth]{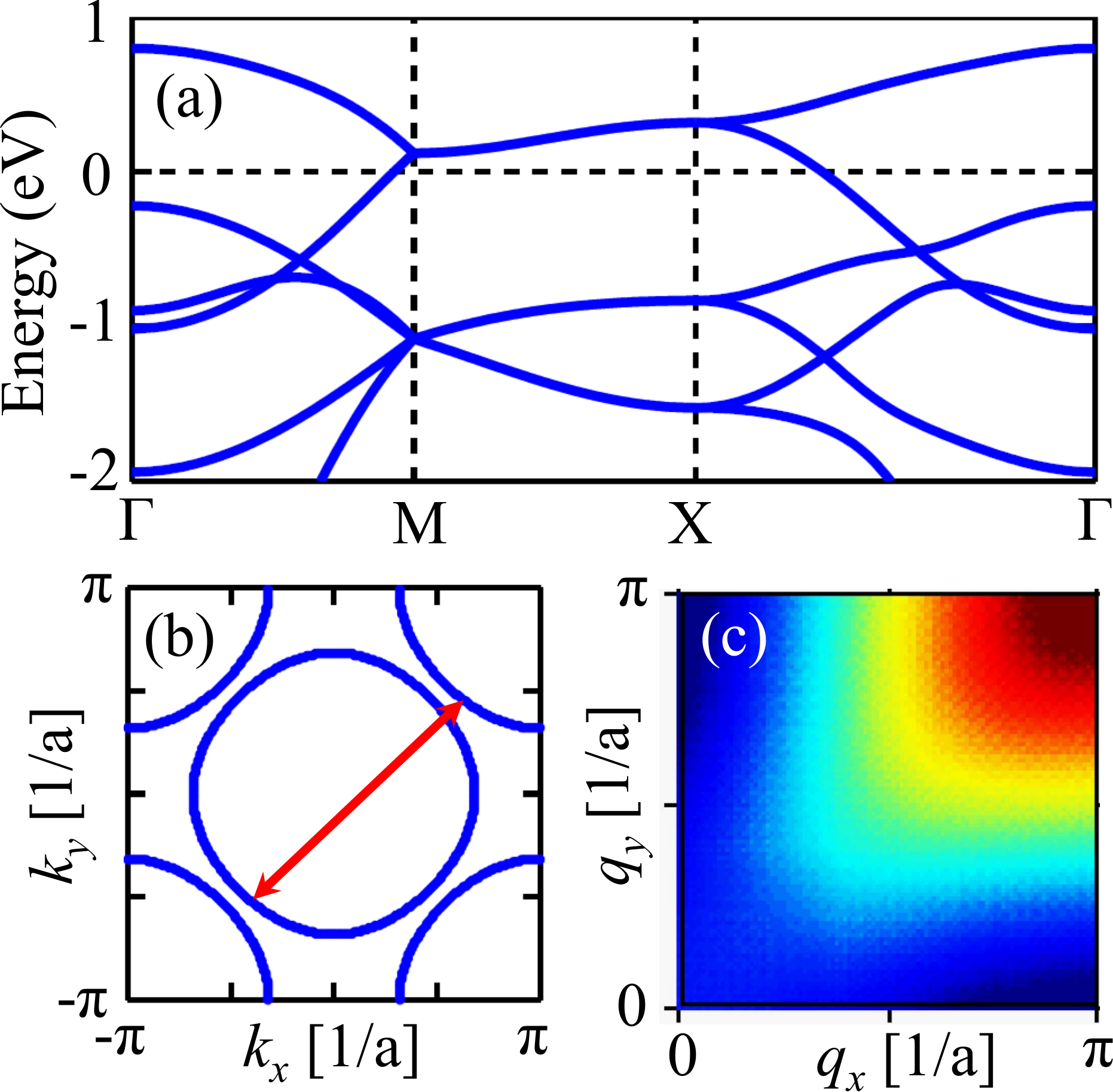}}
 \caption{(Color online) (a) Noninteracting band structure of Sr$_2$IrO$_4$ for SOC $\lambda=0.8$~eV and chemical potential $\mu=0.68$~eV (the other band parameters are same as Ref.~\onlinecite{TB}). Here M=$(\pi/2,\pi/2)$ and X=$(\pi,0)$. (b) Corresponding FS topology. (c) Static in-plane susceptibility showing nesting feature for this material. Red arrow dictates the leading nesting direction between different bands.} \label{fig1}
\end{figure}

Using a three-orbital tight-binding model relevant for Sr$_2$IrO$_4$~\cite{TB}, we study the possible insulating states characterized by various spin, orbital, and spin-orbital entangled density waves by using the RG approach. The materials of interest exhibit antiferromagnetic order above $T^*$ (and below a higher temperature $T_{AF}$ for the magnetic transition). All density waves are anticipated to have an ordering vector $\bf Q=(\pi,\pi)$ in the model of Ref.~\cite{TB} (any system with a dominant nesting vector would have  similar behavior). Also, as this material is not known to be superconducting, we restrict our analysis to particle-hole instabilities. Order parameters then correspond to the condensation of fermion bilinears at momenta ${\bf k}$ and ${\bf k}+{\bf Q}$ in various spin and orbital sectors. In the most general case of SODW,  both the spin and orbital degrees of freedom are different in the condensed fermionic bilinear. Two other cases are spin density wave (SDW) with the same orbital and different spin, and  orbital density wave (ODW) with the same spin and different orbitals.

Following Ref.~\onlinecite{TB}, we use a three $t_{2g}$ orbitals model with SOC, relevant for the Sr$_2$IrO$_4$, we can write the following noninteracting spin-orbit-coupled Hamiltonian:
\begin{multline}
H_0=\sum_{\bm{k}\sigma}\Psi_\sigma^\dagger({\bf k})
\begin{pmatrix}
\xi_1(\bf {k})-\mu & \mathrm{i}\sigma\lambda/2 & -\sigma\lambda/2 \\
-\mathrm{i}\sigma\lambda/2 & \xi_2(\bf{k})-\mu & \mathrm{i}\lambda/2 \\
-\sigma\lambda/2 & -\mathrm{i}\lambda/2 & \xi_3(\bf{k})-\mu
\end{pmatrix}
\Psi_\sigma({\bf k}),
\end{multline}
where $\Psi_\sigma^\dagger({\bf k})\equiv \left(c^{\dagger}_{\bm{k},1\sigma},c^{\dagger}_{\bm{k,}2\sigma},c^{\dagger}_{\bm{k}3,\bar{\sigma}}\right)$, $c_{{\bf k},i\sigma}$ is the annihilation operator for an electron with momentum ${\bf k}$ and spin $\sigma$, residing on orbital $i=1-3$  (respectively the three $t_{2g}$ orbitals $yz$, $xz$, and $xy$), and $\bar{\sigma}\equiv -\sigma$. Here $\xi_{i}({\bf k})$ are the bare dispersion for different orbitals (defined in Ref.~\cite{TB}), and $\lambda$ is the SOC strength and $\mu$ is the chemical potential. 
%As in Ref.~\cite{TB}, we anticipate ordering at  ${\bf Q}=(\pi,\pi)$. 
The eigenvalues of the noninteracting Hamiltonian are then plotted in Fig.~\ref{fig1}(a) and the FS is shown in Fig.~\ref{fig1}(b). The nesting with wavevector $\bf Q$ leads to a logarithmic divergence in the static magnetic susceptibility around this wave vector as shown in Fig.~\ref{fig1}(c). 
%While we focus mainly on the iridate system, our theory of competing symmetry-breaking orders is generic to any system with a %similar FS topology, strong SOC, and a dominant nesting wave vector. With this motivation, we will vary the coupling $\lambda$ as %well as the chemical potential from the values corresponding to Sr$_2$IrO$_4$, and analyze the dependence of the leading %instability on these parameters.

Various orderings can be studied by adding infinitesimal auxiliary fields for SDW, SODW, and ODW [see Fig.~1(b) of the Supplemental Material] to the Hamiltonian
\begin{eqnarray}
&&\Delta_i^{\rm SDW}\sum_{{\bf k},m,n}c^{\dag}_{{\bf k},im}{\sigma}^z_{mn}c_{{\bf k}+{\bf Q},in},\nonumber\\
% \quad
&&\Delta_{ij}^{\rm SODW}\sum_{{\bf k},m, n}c^{\dag}_{{\bf k},im}i\tau^y_{ij}{\bf \sigma}^x_{mn}c_{{\bf k}+{\bf Q},jn},\nonumber\\
&&\Delta_{ij}^{\rm ODW}\sum_{{\bf k},m}c^{\dag}_{{\bf k},im}{\bf \tau}^y_{ij}c_{{\bf k}+{\bf Q},jm}.
\label{fields}
\end{eqnarray}
%
%Constrained by the SOC, SODW occurs between $xz$- and $yz$-orbitals, and ODW commences between $xz$- or $yz$- with $xy$-orbitals, while SDW is allowed for all three orbitals.
Here ${\bm \sigma}$, and ${\bm \tau}$ are the Pauli matrices in the spin-, and orbital-subspaces, and $i$ and $j$ ($m$ and $n$) are orbital (spin)  indices. The most general set of order parameters is rather large so, motivated by experimental results, we have limited the analysis to order parameters given in Eq.~\eqref{fields}. We should note that in this RG treatment, the flow of each order parameter only depends on the flow of the interaction coupling constants [to be deduced in Eqs.~(\ref{sceqn}) below]. Therefore, inclusion and/or exclusion of any order parameter will not affect the flow equations for others.  Experiments  in Sr-based iridates have demonstrated that the spin ordering is collinear with moments oriented along the crystal-axis \cite{neutronSr327,neutron_single_bilayer}. Therefore, we fix the spin orientation to be along the $z$ axis. For the SODW, the symmetry thus allows an associated orbital ordering oriented in the in-plane, i.e. $\tau^x$. We note that a SDW can arise in every orbital, while for the above Hamiltonian, the SODW can naturally form only between $xz$- (orbital index 1) and $yz$- (orbital index 2) orbitals (the spin flip between these orbitals is connected by the nesting vector ${\bm Q}$). Similarly, the non-spin-flip ordering $\Delta^{\rm ODW}$ can commence between the $xy$ with $xz$ and/or $yz$.

We consider all general interactions which include intra- and inter-orbital Coulomb interactions, $U$ and $V$, as well as Hund's coupling $J$, and pair-exchange term $J^{\prime}$ as
\begin{widetext}
{\allowdisplaybreaks\begin{eqnarray}
H_{int}&=&\sum_{{\bf k}_1-{\bf k}_4}\Biggl\{\sum_i U_{i} c^{\dag}_{{\bf k}_1,i\uparrow}c_{{\bf k}_2,i\uparrow}c^{\dag}_{{\bf k}_3,i\downarrow}c_{{\bf k}_4,i\downarrow}+\sum_{i<j,\sigma}\left[V_{ij}c^{\dag}_{{\bf k}_1,i\sigma}c_{{\bf k}_2,i\sigma}c^{\dag}_{{\bf k}_3,j\bar{\sigma}}c_{{\bf k}_4,j\bar{\sigma}}\right.+\left.(V-J)_{ij}c^{\dag}_{{\bf k}_1,i\sigma}c_{{\bf k}_2,i\sigma}c^{\dag}_{{\bf k}_3,j\sigma}c_{{\bf k}_4,j\sigma}\right]\nonumber\\
&&+\sum_{i<j,\sigma}\left(J_{ij}c^{\dag}_{{\bf k}_1,i\sigma}c^{\dag}_{{\bf k}_3,j\bar{\sigma}}c_{{\bf k}_2,i\bar{\sigma}}c_{{\bf k}_4,j\sigma}\right.+\left. J^{\prime}_{ij} c^{\dag}_{{\bf k}_1,i\sigma}c^{\dag}_{{\bf k}_3,i\bar{\sigma}}c_{{\bf k}_2,j\bar{\sigma}}c_{{\bf k}_4,j\sigma} + {\rm H.c.}\right)\Biggr\},
\label{intH}
\end{eqnarray}}
\end{widetext}
subject to the momentum conservation constraint ${\bm k}_1+{\bm k}_3={\bm k}_2+{\bm k}_4$. 
%These interactions terms are shown diagramically in supplementary Fig.~1(c).

{\it Self-consistent order parameters.} As shown in Fig.~\ref{fig1}(b), only two bands contribute to the FS in the realistic range of $\lambda$ and chemical potential. If we define the quasiparticles corresponding to these bands by $\alpha_{{\bm k}}$ and  $\beta_{{\bm k}}$, we can express the orbital in the band basis as
\begin{equation}
c_{{\bm k},i\sigma}\simeq\eta_{{\bm k},i}\alpha_{{\bm k}},\qquad
c_{{\bm k}+{\bm Q},i\sigma}\simeq\gamma_{{\bm k},i}\beta_{{\bm k}},
\label{eigenbasis}
\end{equation}
where the bands deep in the Fermi sea (which do not affect the low-energy physics) have been neglected and $\eta$ and $\gamma$ are the corresponding eigenvectors. Note that these bands are not spin-polarized throughout the Brillouin zone.

The linearized equations for the order parameters are presented graphically in Fig.~1 of the Supplemental Material, which leads to the corresponding self-consistent equations as $1=-T^{n}_{ij}\Gamma^n_{ij}\chi({\bf Q},\Omega)$ where $n$=`SDW', `SODW', and `ODW', and $T^n_{ij}$ are corresponding critical temperatures below which $\Gamma^n_{ij}$ diverges. For a given order, we consider all possible contractions of the interaction term written in Eq.~(\ref{intH}), which leads to the following vertices
\begin{eqnarray}
&&\Gamma^{SDW}_{ii} = \bar{U}_{i} + \sum_{j\ne i} \bar{J}_{ij},~~~{\rm for}~i=1,2,3;\nonumber\\
&&\Gamma^{SODW}_{12} = \bar{V}_{12}+ \bar{J}^{\prime}_{12};\nonumber\\
&&\Gamma^{ODW}_{i3} = (\bar{V}-\bar{J})_{i3} - \bar{J}_{i3}\mp \bar{J}^{\prime}_{i3},~~~{\rm for}~i=1,2.
%
%&&\Gamma^{CDW}_{ii} = \bar{U}_{i} +\sum_{j\neq i} (\bar{V}-\bar{J})_{ij} .
\label{sceqn}
\end{eqnarray}
Here the projected (onto the band basis) interactions terms are {\allowdisplaybreaks $\bar{U}_{i  } = U_{i}\left\langle \eta_{{\bm k}_1,i}\eta_{{\bm k}_2,i}\gamma_{{\bm k}_3,i}\gamma_{{\bm k}_4,i}\right\rangle_{\rm FS}$} ($i$=1,2,3), {\allowdisplaybreaks $\bar{V}_{12}=V_{12}\left\langle \eta_{{\bm k}_1,1}\eta_{{\bm k}_2,1}\gamma_{{\bm k}_3,2}\gamma_{{\bm k}_4,2}\right\rangle_{\rm FS}$}, {\allowdisplaybreaks $(\bar{V}-\bar{J})_{j3}=(V-J)_{j3}\left\langle \eta_{{\bm k}_1,j}\eta_{{\bm k}_2,j}\gamma_{{\bm k}_3,3}\gamma_{{\bm k}_4,3}\right\rangle_{\rm FS}$}, {\allowdisplaybreaks $\bar{J}_{i\ne j}=J_{ij}\left\langle \eta_{{\bm k}_1,i}\gamma_{{\bm k}_3,j}\gamma_{{\bm k}_2,i}\eta_{{\bm k}_4,j}\right\rangle_{\rm FS}$}, and {\allowdisplaybreaks $\bar{J}^{\prime}_{i\ne j}=J^{\prime}_{ij}\left\langle \eta_{{\bm k}_1,i}\eta_{{\bm k}_3,i}\eta_{{\bm k}_2,j}\eta_{{\bm k}_4,j}\right\rangle_{\rm FS}$}, where we have neglected the angle-dependence of the interactions on the FS as a first approximation (momentum-dependent interactions are replaced by their momentum-independent average over the FS). 
We expect this approximation to capture the essential physics given by the topology of the FS and a dominant nesting vector. Using this weak-coupling assumption, we recast the interacting Hamiltonian in Eq.~\eqref{intH} into the band basis as
\begin{eqnarray}
H_{int}&=&\sum_{{\bm k}_1-{\bm k}_4}\left[u_1
\alpha^{\dag}_{{\bm k}_1}\alpha_{{\bm k}_2}\beta^{\dag}_{{\bm k}_3}\beta_{{\bm k}_4}+ u_2 \alpha^{\dag}_{{\bm k}_1}\alpha_{{\bm k}_2}\alpha^{\dag}_{{\bm k}_3}\alpha_{{\bm k}_4}  \right.\nonumber\\
&&~~~~~~+u_3 \beta^{\dag}_{{\bm k}_1}\beta_{{\bm k}_2}\beta^{\dag}_{{\bm k}_3}\beta_{{\bm k}_4}
+ u_4\alpha^{\dag}_{{\bm k}_1}\beta^{\dag}_{{\bm k}_3}\beta_{{\bm k}_2}\alpha_{{\bm k}_4} \nonumber\\
&&~~~~~~\left.+ \left(u_5 \alpha^{\dag}_{{\bm k}_1}\alpha^{\dag}_{{\bm k}_3}\beta_{{\bm k}_2}\beta_{{\bm k}_4}+{\rm H.c.}\right)\right],
\label{eigenintH}
\end{eqnarray}
where interactions $u_{i=1-5}$ are defined as $u_1=\sum_{i=1,3}\bar{U}_{i}+\bar{V}_{12}+\sum_{i=1,2}\left[(\bar{V}-\bar{J})_{i3}-\bar{J}^{\prime}_{i3}\right]$, $u_2=\sum_{i=1,2}\bar{V}_{i3}+(\bar{V}-\bar{J})_{12}$, $u_3=u_2({k}\rightarrow{k}+{Q})$, $u_4=\bar{J}_{12}+\bar{J}^{\prime}_{12}$, $u_5=\sum_{i=1,2}\bar{J}_{i3}$. Note that the complex conjugates of $u_{1,2,3,4}$ are the same as their original forms, and thus not considered separately.

\begin{figure}
%\hspace{-2.2cm}
\rotatebox[origin=c]{0}{\includegraphics[width=0.8\columnwidth]{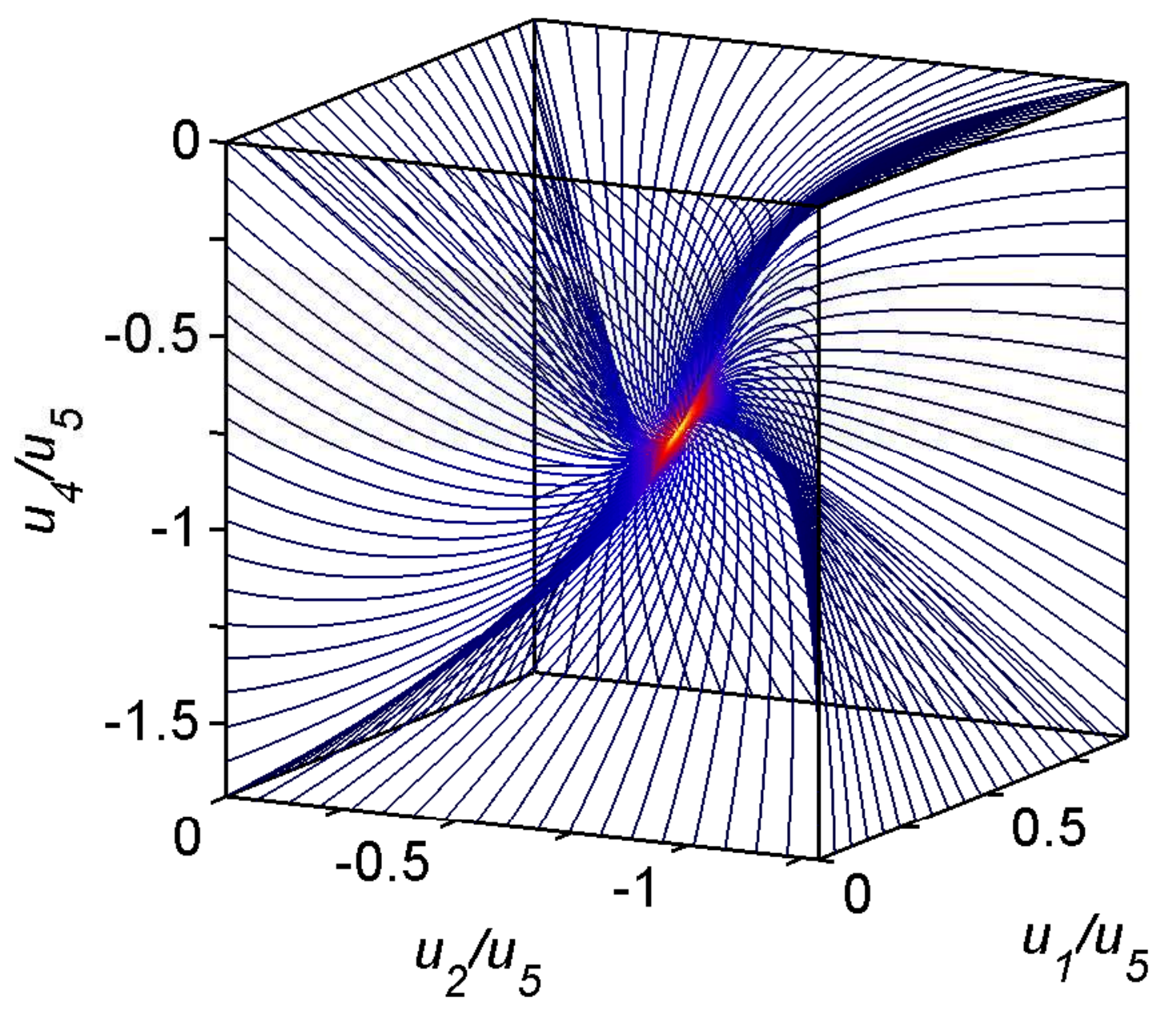}}
 \caption{(Color online) RG flow of Eq.~(\ref{RG}) for various dimensionless coupling constant. As dictated by the equation, $u_2$ and $u_3$ are same at all point in the RG diagram. A gradient color scale is used to highlight the location of the fixed point.} \label{fig2}
\end{figure}
 The derivation of the RG equations is standard as depicted in Fig.~2 of the Supplemental Material, and the corresponding coupled differential equations are
\begin{eqnarray}
\dot{u}_1 &=& 2(u_1^2+u_5u_5^*),\quad
\dot{u}_{2,3}= -2(u_{2,3}^2+u_5u_5^*),\nonumber\\
\dot{u}_{4} &=& 2u_4(u_1+u_4) - 4u_5u_5^*,\quad
\dot{u}_{5} = 2u_5(2u_1-u_2-u_3),
\label{RG}
\end{eqnarray}
where the `dot' symbol represents differentiation with respect to $\ell=1/2 \log{(W/E)}$. The above equations have run-away flows. However, if $u_5\neq 0$, we can recast the above RG equation in terms of the dimensionless coupling constants $u_{1-4}/u_5$ and obtain (both analytically and numerically) stable fixed points for these dimensionless ratios. Our numerical results for the full RG flow in Fig.~\ref{fig2} shows a single stable fixed point for all interactions at $u_1/u_5=0.455$, $u_{2/3}/u_5=-0.644$, and $u_4/u_5=-0.843$. We find that $u_{2/3}$ and $u_4$ obtain finite fixed values only when their initial values are attractive, whereas $u_1$ can be started from any values and reach the same fixed point.

\begin{figure}
%\hspace{-2.2cm}
\rotatebox[origin=c]{0}{\includegraphics[width=0.99\columnwidth]{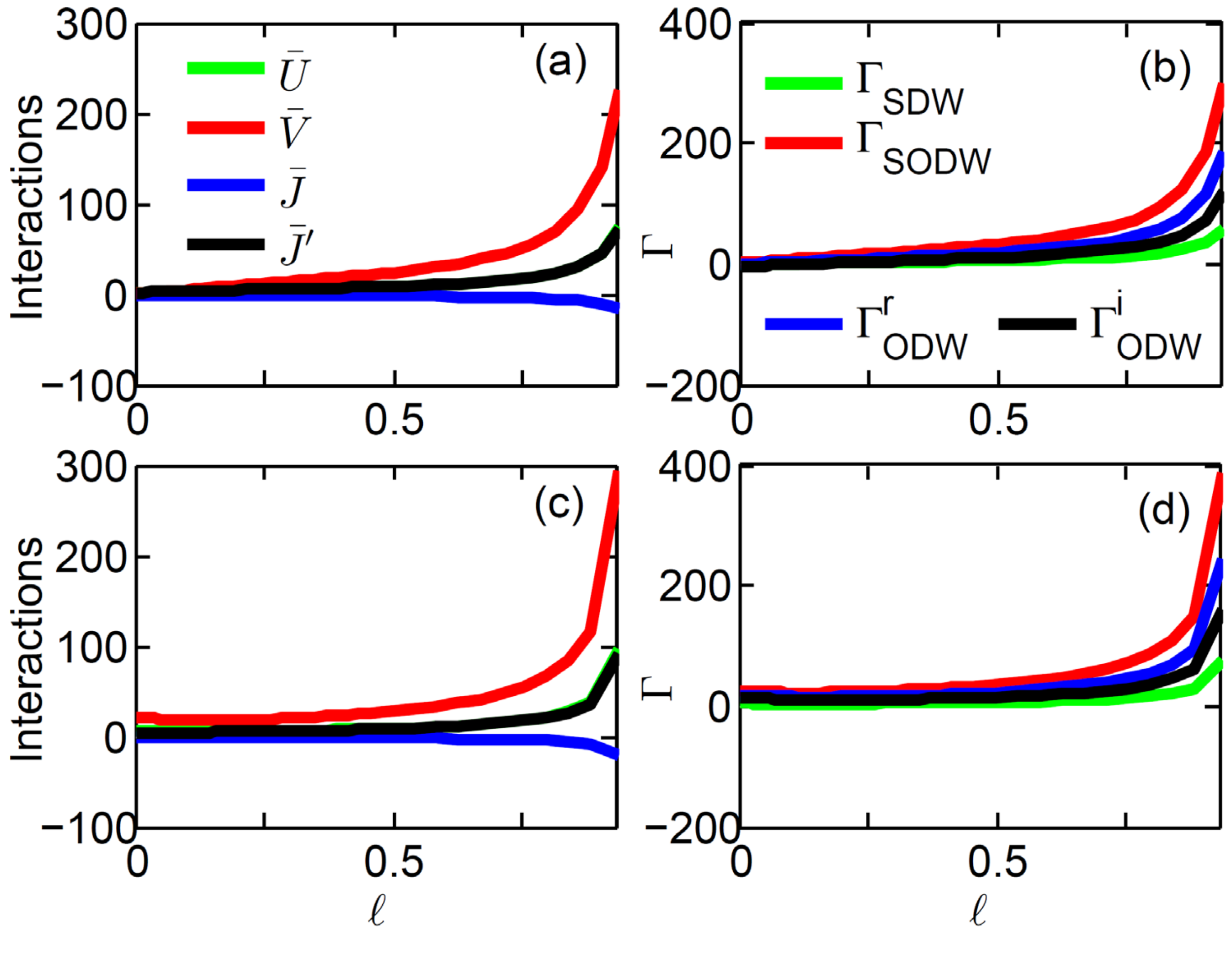}}
\caption{(Color online) (a) RG flow of projected interactions in the orbital basis as given below Eq.~(\ref{sceqn}). (b) Corresponding RG flow of various order parameter vertices. The staring values for all $u^0_i=0$. (c-d) Same as (a) and (b), respectively, but for the starting point of $u_1^0=1$, and $u_{2-4}^0=-1$. For all cases, we find that the inter-orbital coupling $\bar{V}$ dominates which makes the $\Gamma^{SODW}$ term wins over others.
 } \label{fig3}
\end{figure}

Next, we consider the flow of the order-parameter vertices of Eq.~(\ref{sceqn}), and study the dominant divergent density wave at the fixed point in Fig.~\ref{fig3}. Since the flow of the $u_5$ coupling is nontrivial and depends on its bare value, it is appropriate to assume a random-phase approximation (RPA) type flow of this interaction in response to the logarithmic susceptibility as $u_5=u_5^0(1-u_5^0/2 \ell)$, where we have chosen $u_5^0=2$. All the results scales with the initial value of $u_5$ and thus its particular value is irrelevant. 
%At any given chemical potential $\mu$ and SOC $\lambda$, we evaluate the eigenvector of the Hamiltonian and then use unitary transformation of the coupling constant from the band basis ($u_i$) to the orbital basis ($\bar{U}$, $\bar{V}$, $\bar{J}$, $\bar{J}^{\prime}$) following the formulas given below Eq.~(\ref{sceqn}). 
We find that for a wide range of parameters (chemical potential $\mu$ and SOC $\lambda$,) when the two FSs, split by the SOC, have strong interband nesting, the dressed interorbital Coulomb interaction $\bar{V}$ diverges more strongly than other parameters [see  Figs.~\ref{fig3}(a) and \ref{fig3}(c)]. As a result, the SODW dominates over the SDW in such FS topology as shown in Figs.~\ref{fig3}(b) and ~\ref{fig3}(d). 

%Only when the SOC is sufficiently weak, the SDW can become competing [Fig.~\ref{fig3}(a)], however, since both the intraorbital Coulomb term $\bar{U}$ and Hund's coupling $\bar{J}$ are significantly reduced compared to $\bar{U}$ which results in a strong suppression of the SDW in this system.    
%  

\begin{figure}
%\hspace{-2.2cm}
\rotatebox[origin=c]{0}{\includegraphics[width=0.95\columnwidth]{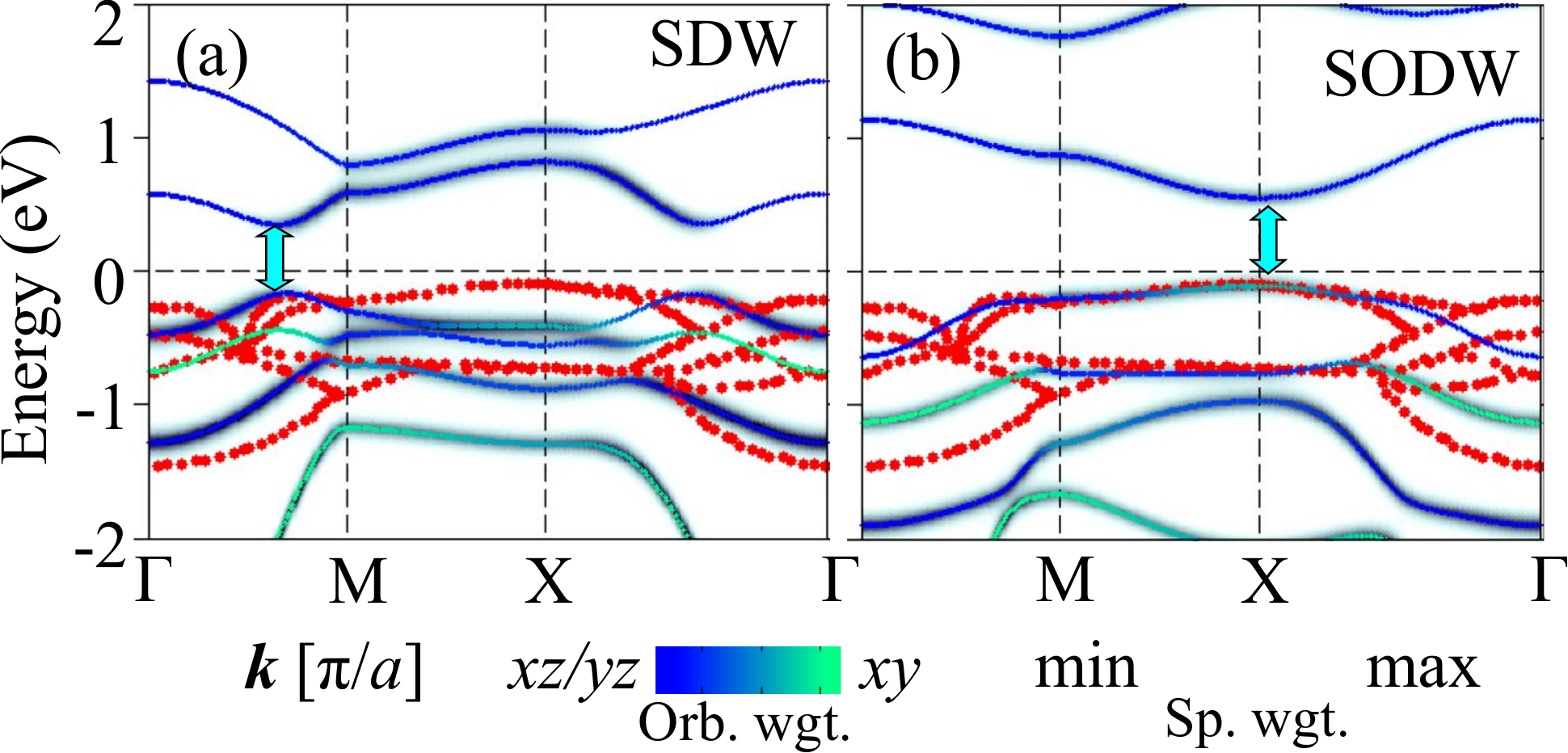}}
\caption{(Color online) Computed spectral weight maps of the insulating Iridates in the SDW (left hand side) and SODW (right hand side) states. The blue to green thin lines highlight the corresponding orbital characters associated with the underlying bands. Red dots are the ARPES data for Sr$_3$Ir$_2$O$_7$ obtained by fitting the maximum of the second derivative of the energy-distribution curves (EDCs) in the insulating state \cite{ARPES}. } \label{fig4}
\end{figure}

Finally, we check our prediction against experimental results on the electronic structure (by employing the mean-field values of the order parameters). Angle-resolved photoemission spectroscopy (ARPES) data \cite{ARPES} in the insulating state of Sr-based iridates pointed out that the direct insulating band gap minimum occurs at the X point, as opposed to at the M points as predicted by LDA+U calculations \cite{LDAU}. A recent scanning tunneling microscopy/spectroscopy (STM/S) also finds inconsistency in the details of the density of states (DOS) with the theoretical calculations \cite{STM}. We find that the ARPES data can be fit well with a SODW gap, rather than any SDW gap. For SDW and SODW, the insulating state can be achieved for $\Delta^{\rm SDW}$=300~meV, and  $\Delta^{\rm SODW}$=1.2~eV with $\lambda=$0.75~eV as shown in the Fig.~\ref{fig4}, which gives the experimental insulating gap for Sr-iridates  \cite{ARPES,STM}. However, for SDW, the band top and bottom on both sides of the insulating gap occurs around the M point, whereas that for the SODW occurs at X point in good agreement with ARPES experiment \cite{ARPES} (plotted by red dots in Fig. 4). On the other hand, for any reasonably large values of the ODW, we do not find an insulting state. This is  because ODW mainly occurs between the $xy$ and $xz/yz$ orbitals and since $xy$ orbital is not present near the Fermi level, ODW does not gap out the FS.

%The SODW may be responsible for a second phase transition (out of an intermediate temperature antiferromagnet) observed in many spin-orbit coupled systems at low temperatures. 
Further evidence of SODW in the experimental data of iridates can be noted. In the bilayer Sr$_3$Ir$_2$O$_7$, e.g., a clear magnetic order sets in below $T_{AF}$=280~K with an increasing DC magnetization, which exhibits a surprising drop below $T^*\sim$70~K \cite{neutronSr327}. More interestingly, the $ab$-plane resistivity shows a power-law temperature dependence below $T_{AF}$ (a behavior expected of a semi-metallic phase), which then switches to an  exponential temperature dependence below $T^*$, marking the onset of the MIT. A similar $T$-dependence of the resistivity and associated magnetic susceptibility is observed in Nd$_2$Ir$_2$O$_7$ with $T_{AF}$=120~K, and $T^*$=8~K \cite{Nd227}. Two magnetic transitions were also found by $\mu$SR in Ca$_5$Ir$_3$O$_{12}$ at $T\sim$300K and 7.8K, in the single layered Sr$_2$IrO$_4$ at $T\sim$100K and 20K \cite{muon}, and in bulk Sr$_2$IrO$_4$ at $T$=100~K and 25~K \cite{MGe}.
%
%
%, with the coexistence of two well-defined precision frequencies.\cite{muon} 
%two magnetization transitions, orientated along the $ab$-plane and $c$-axis, were also observed at $T$=100~K and 25~K in Sr$_2$IrO$_4$ \cite{MGe}.
%, which are below the Curie-Weiss temperature of 200 K in the same sample
% \cite{MGe}.
The onset of MIT after a second magnetic transition (with evidence of a complex electronic texture \cite{neutronSr327}) suggests SODW as a viable candidate for these low-temperature phases. Our RG analysis supports this scenario.

%  
%  These results clearly that not a simple spin-ordering that causes the MIT in this system, but a spin-orbital collective ordering (with zero magnetic moment) drives this transition.

In conclusion, we performed an RG analysis of various possible density-wave orders in systems with strong SOC and a two-band nested FS.  We found that RG flow (toward the stable fixed point) in such systems strongly enhances the relevant inter-orbital coupling interaction, $V$, in comparison to the intra-orbital $U$ or Hund's coupling $J$. We studied the flow of the density-wave vertices for  SDW, SODW, and ODW. We found that the SODW involves $V$, while SDW vertex involves $U$. Therefore, SODW is the dominant order parameter for the above-mentioned FS instability. Furthermore, exploring the role of SODW in MIT, we found that the resulting electronic states of the SODW in Sr$_3$Ir$_2$O$_7$ band structure are in better agreement (than other candidate order parameters) with experimental ARPES results \cite{ARPES}.

\begin{acknowledgments}
We thank Andrey Chubukov for many stimulating discussions and a critical reading our the manuscript. We also thank Cristian Batista and Oskar Vafek for helpful discussions. The work is supported by the U.S. DOE through the Office of Science (BES) and the LANL/LDRD Program and facilitated by NERSC computing allocation.
\end{acknowledgments}

\begin{widetext}
{\Large Supplementary Material}

\section{Self-consistent order parameters}

The order parameters and the interaction Hamiltonian in Eqs.~(2) and (3) of the main text are defined in the orbital basis. Here, we have paid particular attention to the SOC in the noninteracting Hamiltonian which gives, say, spin up for the first two orbitals ($xz$, $yz$) and spin down for the $xy$ orbital at the same momentum. This is, e.g.,  the reason there are only three terms for the interaction $V$ in Fig.~\ref{fig5}(c) of the main text. 
%which corresponds to interactions between different  orbitals with opposite spins and so on. 
For our interaction Hamiltonian, the order parameter vertices are written by considering all possible contractions in the particle-hole channel. In this orbital basis, the self-consistent vertices for each order parameters is graphically expressed in Fig.~\ref{fig5}(b) of the main text. A transformation from the orbital basis to the band basis [using the eigenvectors given in Eq.~(4) of the main text], gives

{\allowdisplaybreaks\begin{eqnarray}
\Gamma^{SDW}_{ii}(\Omega) &=& \left(U_{i}\eta_{{\bm k}_1,i}\eta_{{\bm k}_2,i}\gamma_{{\bm k}_3,i}\gamma_{{\bm k}_4,i}  + \sum_{j\ne i} J_{ij}\eta_{{\bm k}_1,i}\gamma_{{\bm k}_3,j}\gamma_{{\bm k}_2,i}\eta_{{\bm k}_4,j}\right),\qquad i=1,2,3;\nonumber\\
\Gamma^{SODW}_{12}(\Omega) &=& \left(V_{12}\eta_{{\bm k}_1,1}\eta_{{\bm k}_2,1}\gamma_{{\bm k}_3,2}\gamma_{{\bm k}_4,2} + J^{\prime}_{12}\eta_{{\bm k}_1,1}\gamma_{{\bm k}_3,1}\gamma_{{\bm k}_2,2}\eta_{{\bm k}_4,2}\right);\nonumber\\
\Gamma^{ODW}_{i3}(\Omega) &=& \left[(V-J)_{i3}\eta_{{\bm k}_1,i}\eta_{{\bm k}_2,i}\gamma_{{\bm k}_3,3}\gamma_{{\bm k}_4,3}- J_{i3}\eta_{{\bm k}_1,i}\eta_{{\bm k}_3,3}\eta_{{\bm k}_2,i}\eta_{{\bm k}_4,3} \mp J^{\prime}_{i3}\eta_{{\bm k}_1,i}\gamma_{{\bm k}_3,i}\eta_{{\bm k}_2,3}\gamma_{{\bm k}_4,3}\right],~~ i=1,2.
\label{sceqn}
\end{eqnarray}}
With the definitions of various dressed interactions ${\bar U}$, ${\bar V}$, ${\bar J}$ and ${\bar J}^{\prime}$, we obtain Eq.~(5) in the main text. 

\begin{figure*}[t]
\rotatebox[origin=c]{0}{\includegraphics[width=0.7\columnwidth]{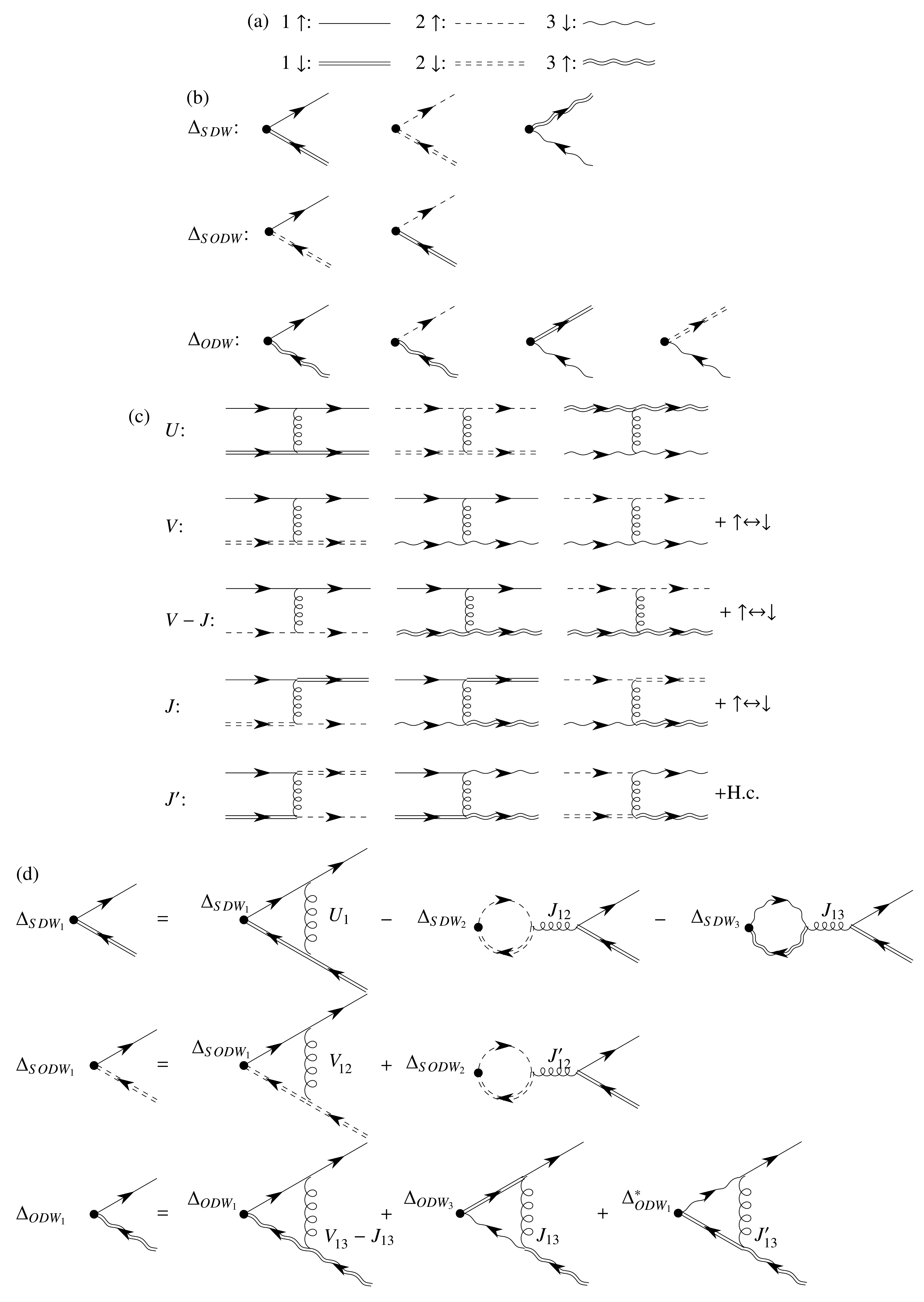}}
\caption{(a) Diagram symbols for spin-polarized orbitals 1$\rightarrow yz$, 2$\rightarrow xz$, and 3$\rightarrow xy$. (b) Various possible order parameters in the particle-hole channel.  (c) The interaction Hamiltonian in Eq.~(3) of the main text, written explicitly for different orbitals. Similar terms obtained for opposite spin follows similarly in which any single line becomes double line, and vice versa. (d) Self-consistent order parameters given in Eqs.~\ref{sceqn}.} \label{fig5}
\end{figure*}

\section{RG equation in the band basis}

Similarly, the interaction Hamiltonian can also be transformed into the band basis as 

%\begin{widetext}
{\allowdisplaybreaks\begin{eqnarray}
H_{I}&=&\sum_{{\bm k}_i}\left[\left(\sum_{i=1,2,3}U_{ii}\eta_{{\bm k}_1,i}\eta_{{\bm k}_2,i}\gamma_{{\bm k}_3,i}\gamma_{{\bm k}_4,i}
+ V_{12}\eta_{{\bm k}_1,1}\eta_{{\bm k}_2,1}\gamma_{{\bm k}_3,2}\gamma_{{\bm k}_4,2} \right.\right. 
\nonumber\\
&&
\left. + \sum_{i=1,2}(V-J)_{i3}\eta_{{\bm k}_1,i}\eta_{{\bm k}_2,i}\gamma_{{\bm k}_3,3}\gamma_{{\bm k}_4,3} -J^{\prime}_{i3}\eta_{{\bm k}_1,i}\gamma_{{\bm k}_3,i}\eta_{{\bm k}_2,3}\gamma_{{\bm k}_4,3}\right)
\alpha^{\dag}_{{\bm k}_1}\alpha_{{\bm k}_2}\beta^{\dag}_{{\bm k}_3}\beta_{{\bm k}_4}
\nonumber\\
&&+\left(\sum_{i=1,2}V_{i3}\eta_{{\bm k}_1,i}\eta_{{\bm k}_2,i}\eta_{{\bm k}_3,3}\eta_{{\bm k}_4,3}
+ (V-J)_{12}\eta_{{\bm k}_1,1}\eta_{{\bm k}_2,1}\eta_{{\bm k}_3,2}\eta_{{\bm k}_4,2}\right)\alpha^{\dag}_{{\bm k}_1}\alpha_{{\bm k}_2}\alpha^{\dag}_{{\bm k}_3}\alpha_{{\bm k}_4}
\nonumber\\
&&+\left(\sum_{i=1,2}V_{i3}\gamma_{{\bm k}_1,i}\gamma_{{\bm k}_2,i}\gamma_{{\bm k}_3,3}\gamma_{{\bm k}_4,3}
+ (V-J)_{12}\gamma_{{\bm k}_1,1}\gamma_{{\bm k}_2,1}\gamma_{{\bm k}_3,2}\gamma_{{\bm k}_4,2}\right)\beta^{\dag}_{{\bm k}_1}\beta_{{\bm k}_2}\beta^{\dag}_{{\bm k}_3}\beta_{{\bm k}_4}\nonumber\\
&&+\left(J_{12}\eta_{{\bm k}_1,1}\gamma_{{\bm k}_3,2}\gamma_{{\bm k}_2,1}\eta_{{\bm k}_4,2}
+ J_{12}^{\prime}\eta_{{\bm k}_1,1}\gamma_{{\bm k}_3,1}\gamma_{{\bm k}_2,2}\eta_{{\bm k}_4,2}\right)\alpha^{\dag}_{{\bm k}_1}\beta^{\dag}_{{\bm k}_3}\beta_{{\bm k}_2}\alpha_{{\bm k}_4}
%\nonumber\\
%
%&&
\left.+\sum_{i=1,2}J_{i3}\eta_{{\bm k}_1,i}\eta_{{\bm k}_3,3}\eta_{{\bm k}_2,i}\eta_{{\bm k}_4,3}\alpha^{\dag}_{{\bm k}_1}\alpha^{\dag}_{{\bm k}_3}\beta_{{\bm k}_2}\beta_{{\bm k}_4}\right]\nonumber\\
&=&\sum_{{\bm k}_i}\left[u_1
\alpha^{\dag}_{{\bm k}_1}\alpha_{{\bm k}_2}\beta^{\dag}_{{\bm k}_3}\beta_{{\bm k}_4}+ u_2 \alpha^{\dag}_{{\bm k}_1}\alpha_{{\bm k}_2}\alpha^{\dag}_{{\bm k}_3}\alpha_{{\bm k}_4} +u_3 \beta^{\dag}_{{\bm k}_1}\beta_{{\bm k}_2}\beta^{\dag}_{{\bm k}_3}\beta_{{\bm k}_4} \right.\left. + u_4\alpha^{\dag}_{{\bm k}_1}\beta^{\dag}_{{\bm k}_3}\beta_{{\bm k}_2}\alpha_{{\bm k}_4} + u_5 \alpha^{\dag}_{{\bm k}_1}\alpha^{\dag}_{{\bm k}_3}\beta_{{\bm k}_2}\beta_{{\bm k}_4}\right].
\label{eigenintH}
\end{eqnarray}}
%\end{widetext}

\begin{figure*}
%\hspace{-2.2cm}
\rotatebox[origin=c]{0}{\includegraphics[width=0.8\columnwidth]{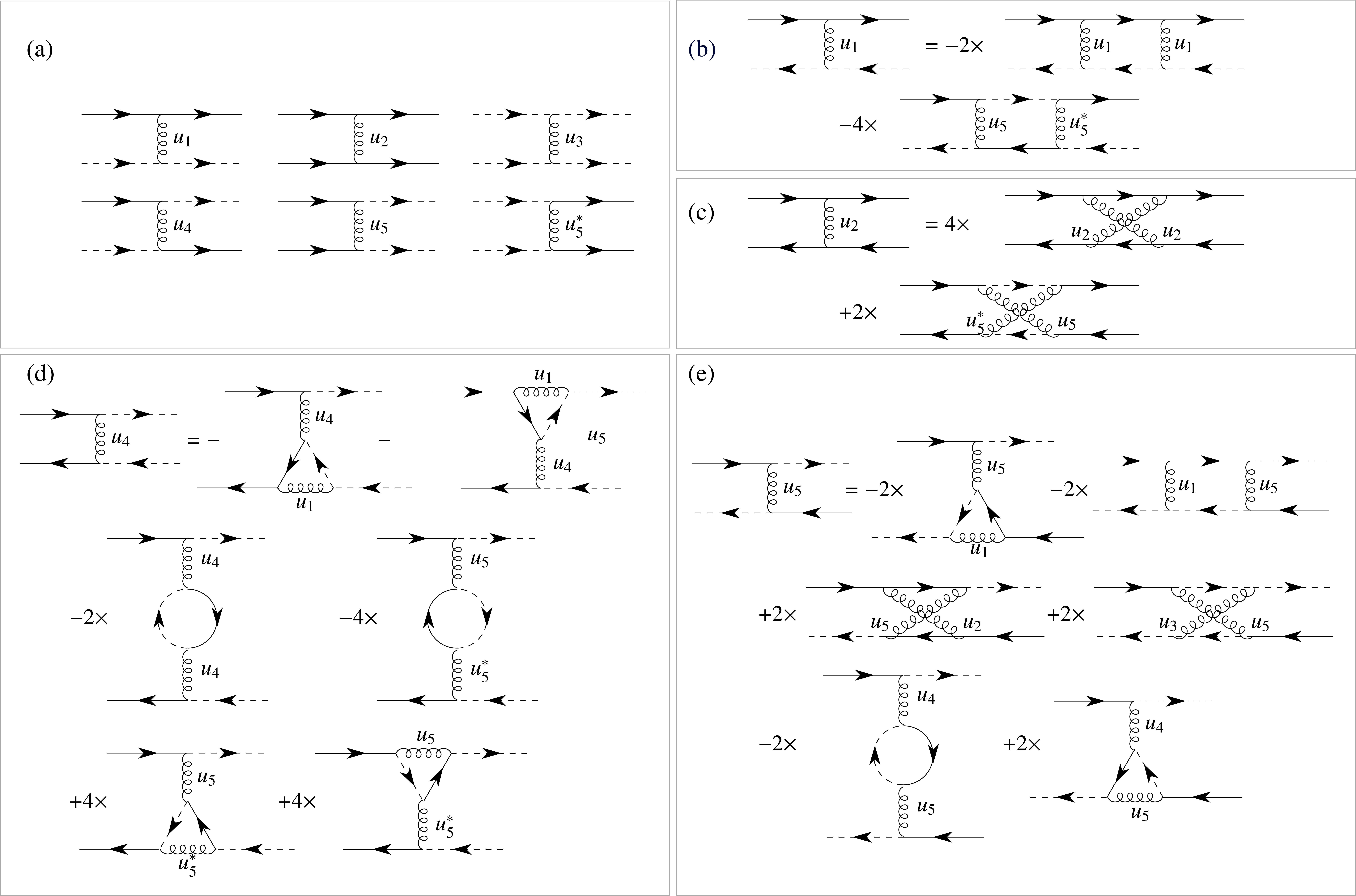}}
 \caption{RG diagrams for the one-loop vertex renormalizations.} \label{fig6}
\end{figure*}

The definition of new interactions $u_{i=1-5}$ in the band basis is readily inferred from the above equation. The new interactions are represented in Fig.~\ref{fig6}(a). The RG equations follow from the diagrams in Fig.~\ref{fig6}, and are given by the following coupled differential equations:
\begin{eqnarray}
{d{u}_1 \over d \ell}= 2(u_1^2+u_5u_5^*),\qquad {d{u}_{2,3}\over d \ell} = -2(u_{2,3}^2+u_5u_5^*),\qquad{d{u}_{4} \over d \ell}&=& 2u_4(u_1+u_4) - 4u_5u_5^*,\qquad {d{u}_{5}\over d \ell} = 2u_5(2u_1-u_2-u_3).
\label{RG}
\end{eqnarray}
Equations above have run-away flows (all $u_i$ diverge at the fixed point). However, the phases of the system are determined by the ratios of these interactions. Assuming $u_5$ is real and nonvanishing, we can formulate the following RG equations for the ratios of the interactions:
{\allowdisplaybreaks\begin{eqnarray}
{d \over d \ell}\left({u_1\over u_5}\right)&=& {2\over u_5}\left(u_5^2-2u_1^2+u_1 u_2+u_1 u_3\right),\qquad {d \over d \ell}\left({u_2\over u_5}\right)= {2\over u_5}\left(-u_5^2-2u_1 u_2 +u_2 u_3\right),\nonumber\\
{d \over d \ell}\left({u_3\over u_5}\right)&=& {2\over u_5}\left(-u_5^2-2u_1 u_3 +u_2 u_3\right),\qquad {d \over d \ell}\left({u_4\over u_5}\right)= {2\over u_5}\left(-2 u_5^2+u_4^2-u_1 u_4 +u_2 u_3+u_2 u_4\right).\nonumber\\
\label{RG2}
\end{eqnarray}}
Setting the right-hand sides of the equations above to zero (corresponding to the fixed-point values), we find that the following fixed-point values are possible: ${u_1\over u_5}=\pm{1\over \sqrt{2\left(1+ \sqrt{2}\right)}}$, ${u_2\over u_5}={u_3\over u_5}=\mp{1\over \sqrt{1+ \sqrt{2}}}$, and ${u_4\over u_5}=\pm{1+\sqrt{2}\pm\sqrt{6\left(2+3\sqrt{2}\right)}\over 2 \sqrt{2\left(1+ \sqrt{2}\right)}}$. The numerical solution of the RG equations, however, indicated hat only one of these possible values correspond to an stable fixed point.

%\begin{figure}
%\hspace{-2.2cm}
%\rotatebox[origin=c]{0}{\includegraphics[width=0.5\columnwidth]{U_flow}}
%\caption{RG flow of various dressed interactions in the orbital space defined in the text.} \label{fig7}
%\end{figure}
%

\end{widetext}

\end{document}